\documentclass[aps,showpacs,11pt]{revtex4}
\usepackage{dcolumn}
\usepackage{graphicx}
\usepackage{amsmath}
\usepackage{amsfonts}
\usepackage{amssymb}
\usepackage{psfrag}
\usepackage{wrapfig}
\usepackage{subfigure}
\usepackage{makeidx}
\usepackage{bm}
\usepackage{epsf}
\usepackage{cases}
\usepackage{hyperref}
\usepackage{multirow}
\makeatletter

\newcommand{\Rmnum}[1]{\expandafter\@slowromancap\romannumeral #1@}
\newcommand{\eq}[1]{Eq.(\ref{#1})}
\makeatother

\begin{document}

\title{Reentrant phase transitions and triple points of topological AdS black holes in Born-Infeld-massive gravity}

\author{Ming Zhang$^{1,2}$\footnote{Corresponding author: zhangming@xaau.edu.cn}, De-Cheng Zou$^{3}$, Rui-hong Yue$^{3}$}

\address{$^{1}$Faculty of Science, Xi'an Aeronautical University, Xi'an 710077 China\\
$^{2}$National Joint Engineering Research Center of Special Pump System Technology, Xi'an 710077, China\\
$^{3}$Center for Gravitation and Cosmology, College of Physical Science and Technology, Yangzhou University, Yangzhou 225009, China}

\date{\today}

\begin{abstract}
\indent

Motivated by recent developments of black hole thermodynamics in de Rham, Gabadadze
and Tolley(dRGT) massive gravity, we study the critical behaviors of
topological Anti-de Sitter(AdS) black holes in the presence of Born-Infeld nonlinear
electrodynamics. Here the cosmological constant appears
as a dynamical pressure of the system and its corresponding
conjugate quantity is interpreted as thermodynamic volume.
It shows that besides the Van der Waals-like SBH/LBH phase transitions,
the so-called reentrant phase transition (RPT) appears in four dimensional
spacetime when the coupling coefficients $c_i m^2$ of massive potential and Born-Infeld
parameter $b$ satisfy some certain conditions. In addition,
we also find the triple critical points and the small/intermediate/large black hole
phase transitions for $d=5$.
\end{abstract}


\maketitle

\section{Introduction}
\label{intro}

The Einstein's General Relativity (GR), which describes the
graviton is a massless spin-2 particle helped us to understand the dynamics
of the Universe \cite{Gupta:1954zz,Weinberg:1965rz,Feynman:1996kb}.
However, there are some fundamental issues, such as the hierarchy problem in particle physics,
the old cosmological constant problem and the origin of late-time acceleration of the Universe
still exist in GR\cite{Capozziello:2011et}.
One of the alternating theory of gravity is known as a massive gravity, where mass
terms are added into the GR action.
A graviton mass has the advantage to potentially provide a theory of dark energy
which could explain the present day acceleration of our Universe \cite{deRham:2014zqa}.
On the other hand, since the quantum theory of massless gravitons is non-renormalizable,
a natural question is whether one can build a self-consistent gravity
theory if the graviton is massive. The first attempt toward constructing
the theory of massive gravity was done by Fierz and Pauli(FP) \cite{Fierz:1939ix}.
With the quadratic order, the FP mass term is the only ghost-free term describing a gravity theory
with five degrees of freedom \cite{VanNieuwenhuizen:1973fi}. However, due to the existence of the
van Dam-Veltman-Zakharov (vDVZ) discontinuity, this theory cannot recover linearized Einstein
gravity in the limit of vanishing graviton mass \cite{vanDam:1970vg,Zakharov:1970cc}.

In particular, Vainshtein \cite{Vainshtein:1972sx} proposed that the linear massive gravity can be recovered
to GR through the 'Vainshtein Mechanism' at small scales by including non-linear terms in the
massive gravity action. Nevertheless, it usually brings various instabilities for the gravitational
theories on the non-linear level by adding generic mass terms, since this model suffers from a pathology
called a 'Boulware-Deser' (BD) ghost.
Later, a new nonlinear massive gravity theory was proposed by
de Rham, Gabadadze and Tolley (dRGT) \cite{deRham:2010ik,deRham:2010kj,Hinterbichler:2011tt},
where the BD ghost \cite{Boulware:1973my} was eliminated by introducing
higher order interaction terms in the action. Then, Vegh \cite{Vegh:2013sk,Adams:2014vza}
constructed a nontrivial black hole solution with a Riccit flat horizon
in four-dimensional dRGT massive gravity. The spherically symmetric solutions were also
addressed in Refs.\cite{Nieuwenhuizen:2011sq,Brito:2013xaa,Ghosh:2015cva}, the corresponding charged
black hole solution was found in \cite{Berezhiani:2011mt,Cai:2014znn}.

Recent development on the thermodynamics of black holes in extended phase space shows that the
cosmological constant can be interpreted as the thermodynamic pressure and treated as
a thermodynamic variable in its own right \cite{Dolan:2011xt,Dolan:2010ha}
\begin{eqnarray}
P=-\frac{\Lambda}{8\pi}\label{PLambda}
\end{eqnarray}
in the geometric units $G_N=\hbar=c=1$. Such operation assume that gravitational
theories including different values of the cosmological constants fall in the ¡°same class¡±, with unified
thermodynamic relations. For black hole thermodynamics, the variation of the cosmological constant
ensures the consistency between the first law of black hole thermodynamics and the Smarr formula.
Moreover, the classical theory of gravity
may be an effective theory which follows from a yet unknown fundamental theory, in which all the
presently 'physical constants' are actually moduli parameters that can run from place to place
in the moduli space of the fundamental theory. Since the fundamental theory is yet unknown, it
is more reasonable to consider the extended thermodynamics of gravitational theories involving
only a single action, and then all variables will appear in the thermodynamical relations.
In the extended phase space, the charged AdS black hole black hole admits a more
direct and precise coincidence between the first order small/large black holes
(SBH/LBH) phase transition and the Van der Waals liquid-gas phase transition,
and both systems share the same critical exponents near the
critical point \cite{Kubiznak:2012wp}. More discussions in various gravity
theories can be found in Refs. \cite{Hansen:2016ayo,Dutta:2013dca,
Xu:2013zea,Dehghani:2014caa,Hendi:2012um,Xu:2014kwa,Cai:2013qga,
Xu:2014tja,Zhao:2014raa,Zou:2014mha,Zhang:2015ova,Zhang:2014eap,Zhang:2014uoa,Majhi:2016txt,Lee:2014tma,
Hendi:2016usw,Kubiznak:2016qmn,Hendi:2016njy,Liu:2014gvf,Kuang:2016caz,Miao:2016ipk,Cadoni:2017ktd}.
Recently, some investigations for thermodynamics of AdS black holes have been
also generalized to the extended phase space in the dRGT massive gravity \cite{Hendi:2017fxp,Xu:2015rfa,
Zhang:2016fxj,Hendi:2015eca}, which show the Van der Waals-like SBH/LBH phase transition
in the charged topological AdS black holes. In addition, the deep relation between the
dynamical perturbation and the Van der Waals-like SBH/LBH phase transition
in the four-dimensional dRGT massive gravity has been also recovered in Ref.\cite{Zou:2017juz}.
In particular, for neutral AdS black holes in all $d\geq6$ dimensional spacetime,
there exist peculiar behavior of intermediate/small/large black hole phase transitions reminiscent of
reentrant phase transitions (RPTs) when the coupling coefficients $c_im^2$ of massive potential
satisfy some certain conditions \cite{Zou:2016sab}. A system undergoes an RPT if a monotonic variation of any
thermodynamic quantity results in two (or more) phase transitions such that the final state is
macroscopically similar to the initial state. The RPT is usually observed in multicomponent fluid systems,
ferroelectrics, gels, liquid crystals, and binary gases \cite{Narayanan}.


In Maxwell's electromagnetic field theory,  a point-like charge which allowed a singularity
at the charge position usually brings about infinite self-energy.
In order to overcome this problem, Born, Infeld \cite{Born:1934gh}
and Hoffmann \cite{Hoffmann:1935ty}
introduced Born-Infeld electromagnetic field to solve infinite self-energy problem by
imposing a maximum strength of the electromagnetic field. In addition,
BI type effective action arises in an open superstring theory and D-branes
are free of physical singularities. In recent two decades, exact solutions of gravitating
black objects in the presence of BI theory have been vastly investigated.
In the extended phase space, Refs.\cite{Gunasekaran:2012dq,Zou:2013owa} recovered the
RPT in the four-dimensional Einstein-Born-Infeld-AdS black hole with spherical horizon.
However, for the higher dimensional Einstein-Born-Infeld AdS black holes
there is no RPT. What about AdS black holes in the
Born-Infeld-massive gravity ? In this paper, we will generalize the discussion
to topological AdS black holes for $d=4$ and 5 in the
Born-Infeld-massive gravity.

This paper is organized as follows. In Sect.~\ref{2s}, we review the
thermodynamics of Born-Infeld-massive black holes in the extended phase space.
In Sect.~\ref{3s}, we study the critical behaviors of four and five dimensional topological
AdS black holes in context of $P-V$ criticality and phase diagrams.
We end the paper with conclusions and discussions in Sect.~\ref{4s}.

\section{Thermodynamics of $d$-dimensional Born-Infeld AdS black holes}
\label{2s}

We start with the action of $d$-dimensional massive gravity in presence of Born-Infeld
field \cite{Hendi:2015hoa}
\begin{eqnarray}
\mathcal{I}=\frac{1}{16\pi}\int{d^{d} x\sqrt{-g}\left[R-2\Lambda+{\cal L(F)}+m^2\sum_{i=1}^{4}c_{i}{\cal U}_{i}(g,f)\right]},\label{action}
\end{eqnarray}
where the last four terms are the massive potential associate with graviton mass,
$c_i$ are the negative constants\cite{Cai:2014znn} and $f$ is a fixed rank-2 symmetric tensor.
Moreover, ${\cal U}_{i}$ are symmetric polynomials of the eigenvalues
of the $d\times d$ matrix ${\cal K}^\mu_{\nu}\equiv\sqrt{g^{\mu\alpha}f_{\alpha\nu}}$
\begin{eqnarray}
{\cal U}_{1}&=&[\cal K],\nonumber\\
{\cal U}_{2}&=&[{\cal K}]^2-[{\cal K}^2],\nonumber\\
{\cal U}_{3}&=&[{\cal K}]^3-3[{\cal K}][{\cal K}^2]+2[{\cal K}^3],\nonumber\\
{\cal U}_{4}&=&[{\cal K}]^4-6[{\cal K}^2][{\cal K}]^2
+8[{\cal K}^3][{\cal K}]
+3[{\cal K}^2]^2-6[{\cal K}^4].\label{ac}
\end{eqnarray}
The square root in ${\cal K}$ is understood as the matrix square root, ie.,
$(\sqrt{A})^{\mu}_{~\nu}(\sqrt{A})^{~\nu}_{\lambda}=A^\mu_{~\lambda}$, and the
rectangular brackets denote traces $[{\cal K}]={\cal K}^{\mu}_{~\mu}$.
In addition, $b$ is the Born-Infeld parameter and ${\cal L(F)}$ with
\begin{eqnarray}
{\cal L(F)}=4b^2\left(1-\sqrt{1+\frac{F^{\mu\nu}F_{\mu\nu}}{2b^2}}\right). \label{eq:2a}
\end{eqnarray}
In the limit $b\rightarrow \infty$, it reduces to the standard Maxwell field ${\cal
L(F)}=-F^{\mu\nu}F_{\mu\nu}+\mathcal {O}(F^4)$. If taking $b=0$, ${\cal L(F)}$ disappears.

Consider the metric of $d$-dimensional spacetime in the following form
\begin{eqnarray}
ds^2=-f(r)dt^2+\frac{1}{f(r)}dr^2+r^2h_{ij}dx^idx^j,\label{metric}
\end{eqnarray}
where $h_{ij}dx^idx^j$ is the line element for
an Einstein space with constant curvature $(d-2)(d-3)k$.
The constant $k$ characterizes the geometric property of
hypersurface, which takes values $k=0$ for flat, $k=-1$ for
negative curvature and $k=1$ for positive curvature, respectively.

By using the reference metric \cite{Cai:2014znn}
\begin{eqnarray}
f_{\mu\nu}=diag(0,0,c_0^2h_{ij})
\end{eqnarray}
with a positive constant $c_0$, we can obtain
\begin{eqnarray}
{\cal U}_{1}&=&(d-2)c_0/r,\nonumber\\
{\cal U}_{2}&=&(d-2)(d-3)c_0^2/r^2,\nonumber\\
{\cal U}_{3}&=&(d-2)(d-3)(d-4)c_0^3/r^3,\nonumber\\
{\cal U}_{4}&=&(d-2)(d-3)(d-4)(d-5)c_0^4/r^4.
\end{eqnarray}
Obviously, the terms related with $c_3$ and $c_4$ only appear
in the black hole solutions for $d\geq5$ and $d\geq6$, respectively \cite{Cai:2014znn}.

In addition, the electromagnetic field tensor in $d$-dimensions is given by
$F_{tr}=\sqrt{\frac{d_2d_3}{\left(1+\Gamma\right)}}\frac{q}{r^{d_2}}$,
and the metric function $f(r)$ is obtained as \cite{Hendi:2015hoa}
\begin{eqnarray}
f(r)&=&k-\frac{m_0}{r^{d_3}}+\frac{\left(4b^2-2\Lambda\right)}{d_1d_2} r^2-\frac{4b^2r^2}{d_1 d_2}\sqrt{1+\Gamma}+\frac{4d_2q^2}{d_1r^{2d_3}}{\cal H}\nonumber\\
&&+m^2c_0\left(\frac{c_1 r}{d_2}+c_0c_2+\frac{d_3c_0^2c_3}{r}
+\frac{d_3d_4c_0^3c_4}{r^2}\right), \label{solution}
\end{eqnarray}
where $d_i=d-i$ and
\begin{eqnarray}
\Gamma=\frac{d_2d_3q^2}{b^2r^{2d_2}},\quad
{\cal H}={_2}F_1[\frac{1}{2},\frac{d_3}{2d_2},\frac{3d_{7/3}}{2d_2},-\Gamma].
\end{eqnarray}
Moreover, $m_0$ and $q$ are related to the mass $M$ and charge $Q$ of black holes as
\begin{eqnarray}
Q=\frac{\sqrt{d_2d_3}\Sigma_k }{4\pi}q, \quad M=\frac{d_2\Sigma_k}{16\pi}m_0,\label{eq:6a}
\end{eqnarray}
where $\Sigma_k$ represents the volume of constant curvature hypersurface described by
$h_{ij}dx^idx^j$. The electromagnetic potential difference ($\Phi$) between the horizon
and infinity reads as $\Phi=\sqrt{\frac{d_2}{d_3}}\frac{q}{r_+^{d_3}}{\cal H_+}$.

Then the mass $M$ of the Born-Infeld AdS black hole for massive gravity is given by
\begin{eqnarray}
M&=&\frac{d_2\Sigma_kr_+^{d_3}}{16\pi}\left[k+\frac{16\pi P}{d_1d_2}r_+^2
+\frac{4b^2r_+^2}{d_1d_2}\left(1-\sqrt{1+\Gamma_+}\right)
+\frac{4d_2 q^2}{d_1r_+^{2d_3}}{\cal H_+}\right.\nonumber\\
&&\left.+m^2\left(\frac{c_0c_1r_+}{d_2}+c_0^2c_2+\frac{d_3c_0^3c_3}{r_+}
+\frac{d_3d_4c_0^4c_4}{r_+^2}\right)\right],\label{Mass}
\end{eqnarray}
in terms of the horizon radius $r_+$. Due to existence of the pressure in obtained relation
for total mass of the black holes, here the black hole mass $M$ can be considered as the
enthalpy $H$ rather than the internal energy of the gravitational system \cite{Kastor:2009wy}.

In addition, the Hawking temperature which is related to the definition of
surface gravity on the outer horizon $r_+$ can be obtained as
\begin{eqnarray}
T&=&\frac{d_3k}{4\pi r_+}+\frac{4r_+}{d_2}P
+\frac{b^2r_+}{d_2\pi}\left(1-\sqrt{1+\Gamma_+}\right)\nonumber\\
&&+\frac{m^2c_0}{4\pi}\left(c_1+\frac{d_3c_0c_2}{r_{+}}+\frac{d_3d_4c_0^2c_3}{r_+^2}
+\frac{d_3d_4d_5c_0^3c_4}{r_+^3}\right),\label{Tem}
\end{eqnarray}
and the entropy $S$ of the Born-Infeld AdS black hole reads as
\begin{eqnarray}
S=\frac{\Sigma_k}{4} r_+^{d_2} \label{entropy}
\end{eqnarray}
It is easy to check that those thermodynamic quantities obey
the (extended phase-space) first law of black hole thermodynamics
\begin{eqnarray}
dH&=&TdS+VdP+\mathfrak{B}db+\frac{c_0m^2\Sigma_k r_+^{d_2}}{16\pi}dc_1
+\frac{d_2c_0^2m^2\Sigma_k r_+^{d_3}}{16\pi}dc_2\nonumber\\
&&+\frac{d_2d_3c_0^3m^2\Sigma_k r_+^{d_4}}{16\pi}dc_3
+\frac{d_2d_3d_4c_0^4m^2\Sigma_k r_+^{d_5}}{16\pi}dc_4\label{flbh},
\end{eqnarray}
where $\mathfrak{B}$, which is a quantity conjugate to $b$ is called
the ``Born-Infeld vacuum polarization''
\begin{eqnarray}
\mathfrak{B}=\left(\frac{\partial H}{\partial b}\right)|_{(S,P,c_1,c_2,c_3,c_4)}=\frac{\Sigma_k b r_+^{d_1}}{2d_1\pi}\left(1-\sqrt{1+\Gamma_+}\right)
+\frac{d_2d_3\Sigma_k}{4\pi d_1b}\frac{{\cal H_+}q^2}{ r_+^{d_3}},\label{B}
\end{eqnarray}
the thermodynamic volume $V$ \cite{Cvetic:2010jb}, which is the corresponding conjugate quantity of $P$, can be written as
\begin{eqnarray}
V=\frac{\Sigma_k r_1^{d_1}}{d_1}.
\end{eqnarray}

The behavior of free energy $G$ is important to determine the thermodynamic phase
transition in the canonical ensemble. We can calculate the free energy from the thermodynamic
relation
\begin{eqnarray}
G=H-TS&=&\frac{r_+^{d_1}}{d_1d_2}\left(P+\frac{b^2}{4\pi}\sqrt{1+\Gamma_+}\right)
+\frac{d_2^2 q^2{\cal H_+}}{2\pi d_1 r_+^{d_3}}+\frac{r_+^{d_3}}{16\pi}\nonumber\\
&&+\frac{m^2c_0^2r_+^{d_5}}{16\pi}\left(c_2r_+^2+2d_3c_0c_3r_++3d_3d_4c_0^2c_4\right).
\end{eqnarray}

\section{Phase transitions of topological AdS black holes in Born-Infeld-Massive gravity}
\label{3s}

For further convenience, we denote
\begin{eqnarray}
&&\hat{T}=T-\frac{c_0c_1m^2}{4\pi},\quad W_2=-\frac{k+c_0^2c_2m^2}{8\pi},\nonumber\\
&&W_3=-\frac{c_0^3c_3m^2}{8\pi},\quad W_4=-\frac{c_0^4c_4m^2}{8\pi},\label{wcoef}
\end{eqnarray}
Here $\hat{T}$ denotes the shifted temperature and can be negative according
to the value of $c_0c_1m^2$.
Then, the equation of state of the black hole can be obtained from Eq.~(\ref{Tem})
\begin{eqnarray}
P&=&\frac{d_2}{4r_+}\left[\hat{T}+\frac{2d_3W_2}{r_+}+\frac{2d_3d_4W_3}{r_+^2}
+\frac{2d_3d_4d_5W_4}{r_+^3}-\frac{b^2r_+}{d_2\pi}\left(1
-\sqrt{1+\Gamma_+}\right)\right].\label{eos}
\end{eqnarray}
To compare with the VdW fluid equation, we can translate
the ``geometric" equation of state to physical one by identifying the specific
volume $v$ of the fluid with the horizon radius of the black hole as $v=\frac{4r_+}{d_2}$.
Evidently, the specific volume $v$ is proportional to the horizon radius $r_+$,
therefore we will just use the horizon radius in the equation of
state for the black hole hereafter in this paper.

We know that the critical point occurs when $P$
has an inflection point,
\begin{eqnarray}
\frac{\partial P}{\partial r_+}\Big|_{\hat{T}=\hat{T}_c, r_+=r_c}
=\frac{\partial^2 P}{\partial r_+^2}\Big|_{\hat{T}=\hat{T}_c, r_+=r_c}=0,
\label{eq:15a}
\end{eqnarray}
where the subscript stands for the quantities at the critical point.
The critical shifted temperature is obtained as
\begin{eqnarray}
\hat{T}_{c}=-\frac{2d_3}{r_c}\left(2w_2+\frac{3d_4W_3}{r_c}
+\frac{4d_4d_5W_4}{r_c^2}\right)-\frac{d_2d_3q^2}{\pi r_c^{2d_{5/2}}}\left(1
+\Gamma_+\right)^{-1/2},\label{Tcrit}
\end{eqnarray}
and the equation for critical horizon radius $r_c$ is given by
\begin{eqnarray}
F(r_c)&=&6d_4d_5W_4+3d_4W_3r_c+W_2r_c^2+\frac{d_{5/2} d_2 q^2}{2\pi r_c^{2d_4}}\left(1
+\Gamma_+\right)^{-1/2}\nonumber\\
&&-\frac{d_3 d_2^3 q^4}{4\pi b^2r_c^{4d_3}}\left(1
+\Gamma_+\right)^{-3/2}=0.\label{crit}
\end{eqnarray}

For later discussions, it is convenient to rescale some quantities in the following way
\begin{eqnarray}
&&W_2=q^{\frac{2}{d-2}}\cdot b^{\frac{2(d-3)}{d-2}}w_2,\quad W_3=q^{\frac{3}{d-2}}\cdot b^{\frac{2d-7}{d-2}}w_3,\quad
W_4=q^{\frac{4}{d-2}}\cdot b^{\frac{2(d-4)}{d-2}}w_4\nonumber\\
&&r_+=\left(\frac{q}{b}\right)^{\frac{1}{d-2}}\cdot x,\quad P=b^2\cdot p,\quad
\hat{T}=q^{\frac{1}{d-2}}\cdot b^{\frac{2d-5}{d-2}}\cdot t,\quad
G=q^{\frac{d-1}{d-2}}\cdot b^{-\frac{2}{d-2}}\Sigma_k\cdot g.
\end{eqnarray}
In terms of quantities above, Eqs.(\ref{eos}), (\ref{Tcrit}) and (\ref{crit})
can be written as
\begin{eqnarray}
p&=&\frac{d_2}{4x}\left[t+\frac{2d_3w_2}{x}+\frac{2d_3d_4w_3}{x^2}
+\frac{2d_3d_4d_5w_4}{x^3}-\frac{x}{d_2\pi}\left(1
-\sqrt{1+\frac{d_2d_3}{x^{2d_2}}}\right)\right],\label{eos1}\\
t_{c}&=&-\frac{2d_3}{x_c}\left(2w_2+\frac{3d_4w_3}{x_c}
+\frac{4d_4d_5w_4}{x_c^2}\right)-\frac{d_2d_3}{\pi x_c^{2d_{5/2}}}\left(1
+\frac{d_2d_3}{x_c^{2d_2}}\right)^{-1/2},\label{Tcrit1}\\
F(x_c)&=&6d_4d_5w_4+3d_4w_3x_c+w_2x_c^2+\frac{ d_2d_{5/2}}{2\pi x_c^{2d_4}}\left(1
+\frac{d_2d_3}{x_c^{2d_2}}\right)^{-1/2}\nonumber\\
&&-\frac{d_3 d_2^3}{4\pi x_c^{4d_3}}\left(1
+\frac{d_2d_3}{x_c^{2d_2}}\right)^{-3/2}=0,\label{crit1}
\end{eqnarray}
where $x_c$ denotes the critical value of $x$.
For arbitrary parameter $d$, it is hard to obtain the exact solution of \eq{crit1}.

In what follows we shall specialize to $d=4$ and 5,
and then perform a detailed study of the thermodynamics of these black holes.

\subsection{$P-V$ criticality for $d=4$}
For $d=4$, \eq{crit1} will reduce to the cubic equation
\begin{eqnarray}
F(y)=y^{3}-\frac{3y}{4}-\frac{\pi w_2}{2}=0\label{Frc}
\end{eqnarray}
with $y=\left(x_c^4+2\right)^{-1/2}$.

Depending on different values of $w_2$, Eq.~(\ref{Frc}) admits one or more positive
real roots for $x$, which can be also reflected by
\begin{eqnarray}
\frac{\partial F(y)}{\partial y}=3y^{2}-\frac{3}{4}.\label{dFrc}
\end{eqnarray}

When $\left\lvert w_2\right\rvert\leq \frac{1}{2\pi}$,
three real roots occur, which are given by
\begin{eqnarray}
y_i=\cos\left(\frac{1}{3}\arccos\left(2\pi w_2\right)-\frac{2\pi i}{3}\right),\quad i=0,1,2.\label{root}
\end{eqnarray}
Moreover, in order that $x_c=\left(\frac{1}{y^2}-2\right)^{1/4}$ be positive,
we require an additional constraint $\left\lvert y\right\rvert\leq \frac{1}{\sqrt{2}}$.
Then, we have $y_0>0$ in case of $-\frac{1}{2\pi}\leq w_2\leq -\frac{1}{\sqrt{8}\pi}$,
and $y_1>0$ in the region of $-\frac{1}{2\pi}\leq w_2\leq 0$,  while the solution $y_2$
is always negative.

Now by inserting solutions of $y_0$ and $y_1$ into Eqs.~(\ref{eos1}) and (\ref{Tcrit1}),
we analyze the critical behaviors. Notice that analytic methods can not be
applied in our analysis because of the complexity of the Gibbs free energy
and equation of state, we resort to graphical and numerical methods.

\begin{enumerate}
\item $w_2\in(-\frac{1}{\sqrt{8}\pi},0)$. As shown in Fig.~{\ref{fig1}},
the $p-x$ diagram displays that the dashed curve represents critical isotherm at $t=t_c$,
the dotted and solid curves correspond to $t>t_c$ and $t<t_c$, respectively.
In the $g-t$ diagram, the solid curve represents $p<p_c$, the dotted curve
correspond to $p>p_c$ and the dashed curve is for $p=p_c$. We observe standard swallowtail
behavior. Moreover, the $p-t$ diagram shows the coexistence line of the first-order
phase transition terminating at a critical point. These plots are analogous to typical
behavior of the liquid-gas phase transition of the Van der Waals fluid.

\begin{figure}[htb]
  \includegraphics[width=0.3\textwidth]{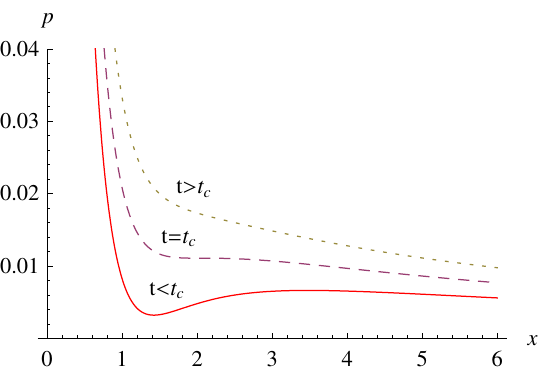}
\hfill%
  \includegraphics[width=0.3\textwidth]{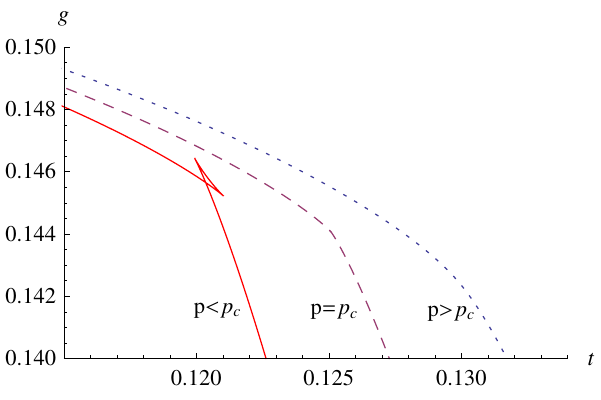}
  \hfill%
  \includegraphics[width=0.3\textwidth]{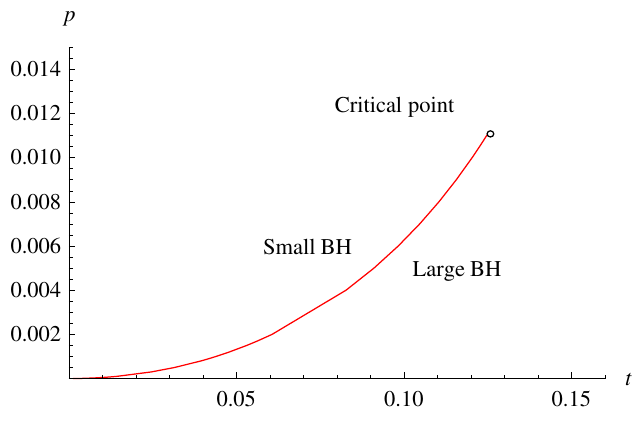}
  \caption{Born-Infeld AdS black holes for $d=4$ and $w_2=-0.1$.
There is one critical point, which corresponds VdW-like SBH/LBH phase transition
when $t<t_c$. Here the critical pressure and temperature read $p=p_c\approx0.011986$
and $t=t_c\approx0.125087$, respectively. }\label{fig1}
\end{figure}

\item $w_2\in(-0.132795,-\frac{1}{\sqrt{8}\pi})$,
there only exist one physical (with positive pressure) critical point and the
corresponding VdW-like SBH/LBH phase transition, which occurs for the pressures
$p\in(p_\tau, p_{c1})$ and temperatures $t\in(t_\tau, t_{c1})$, see Fig.~{\ref{fig2}}.
For the $p-t$ diagram in Fig.~{\ref{fig3}}, three separate phases of black holes
emerge in the region of $p_\tau<p\leq p_z<p_{c1}$: intermediate black holes
(IBH) (on the left), small (on the middle), and large (on the right), where
small and large black holes are separated by
the SBH/LBH phase transition, but the intermediate and small
are separated by a finite jump in $g$, which is so-called \textit{zeroth-order
phase transition}\cite{Maslov}.
For $p<p_{\tau}$ only one phase of large black holes exists.
When taking $w_2=-0.124$, we obtain
\begin{eqnarray}
&&(t_\tau, t_z, t_{c1})\approx(0.12316, 0.123825, 0.175593),\nonumber\\
&&(p_\tau, p_z, p_{c1})\approx(0.0072472, 0.008104, 0.0177545)
\end{eqnarray}

\begin{figure}[htb]
  \includegraphics[width=0.3\textwidth]{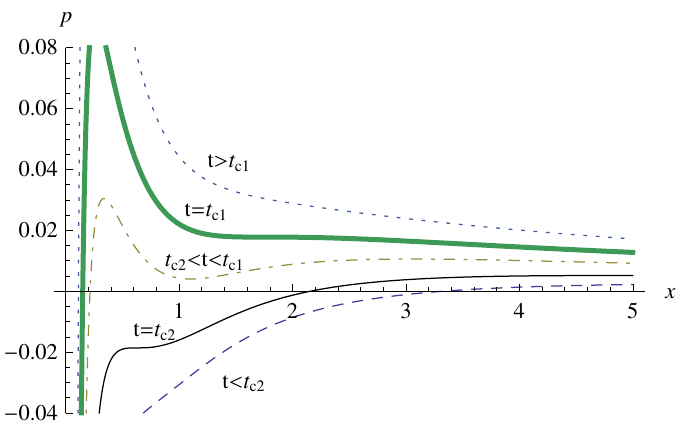}
  \hfill%
  \includegraphics[width=0.3\textwidth]{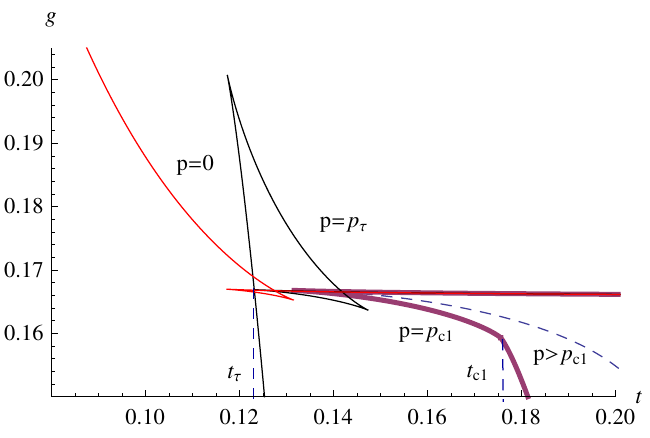}
  \hfill%
  \includegraphics[width=0.3\textwidth]{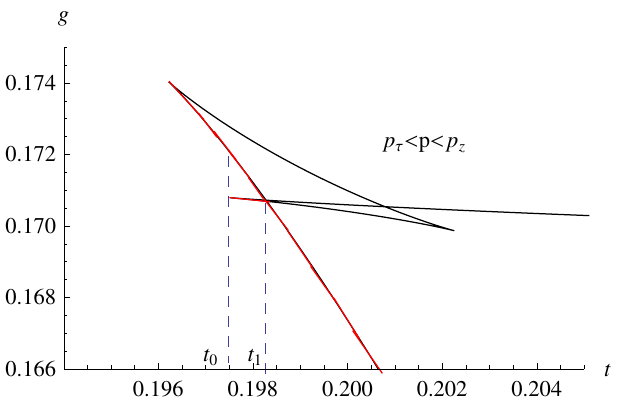}
  \caption{Born-Infeld AdS black holes for $d=4$ and $w_2=-0.124$.
Left: the $p-x$ diagram shows the existence of two critical points,
one at positive pressure $p_{c1}\approx0.0177545$, the
other at negative pressure $p_{c2}\approx-0.0084877$.
Center: the Gibbs free energy shows one physical (with positive pressure)
critical point and the corresponding first order SBH/LBH phase transition,
occurring for $t\in(t_{\tau}, t_{c1})$ and $p\in(p_{\tau}, p_{c1})$.
Right: there is a \textit{reentrant phase transition} (RPT) corresponding to the \textit{zeroth-order
phase transition} at $t=t_0$ followed by a first-order VdW-like SBH/LBH
phase transition at the intersection $t=t_1$ with the swallowtail structure.}\label{fig2}
\end{figure}

\begin{figure}[htb]
\centering
\includegraphics{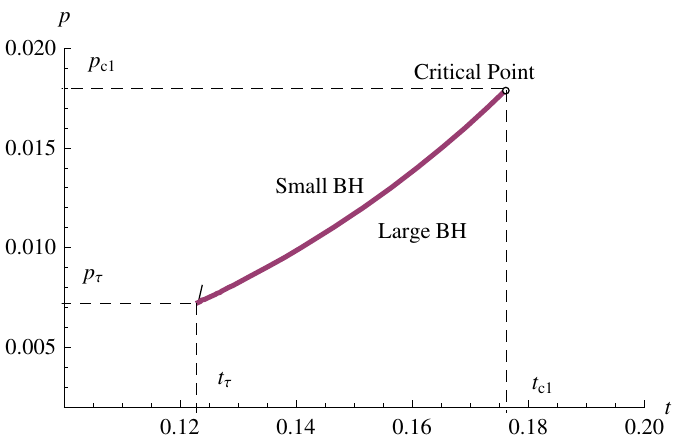}
    \hfill%
\includegraphics[width=0.4\textwidth]{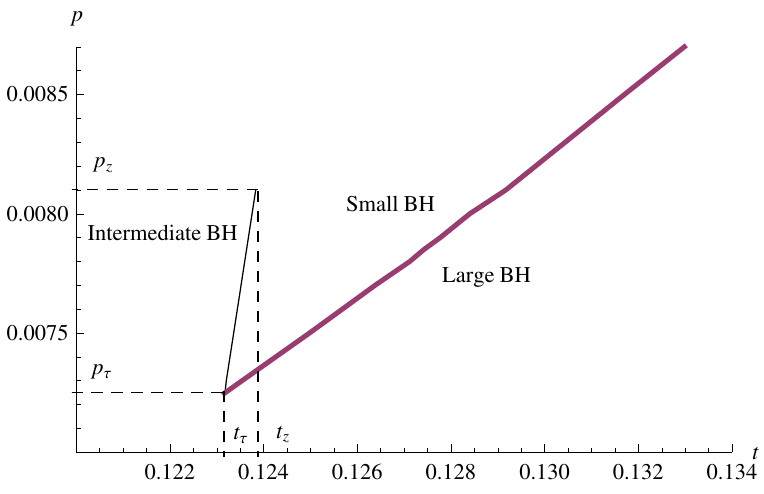}
\caption{The coexistence line of the VdW-like SBH/LBH phase transition
is depicted by a thick solid line. It initiates
from the critical point $(p_{c1}, t_{c1})$ and terminates at $(p_{\tau}, t_{\tau})$.
The solid line describes the `coexistence line' of small
and intermediate black holes, separated by a finite gap in $g$,
indicating the zeroth order phase transition. It commences
from $(t_z, p_z)$ and terminates at $(p_{\tau}, t_{\tau})$.}\label{fig3}
\end{figure}

\item $w_2\in(-\frac{1}{2\pi},-0.132795)$, there exist
two critical points with positive pressure, and the similar RPT
also occurs. As shown in Fig.~{\ref{fig4}}, we obtain
\begin{eqnarray}
&&(t_{c2}, t_\tau, t_z, t_{c1})\approx(0.187113,0.197121, 0.198064, 0.2139999),\nonumber\\
&&(p_{c2}, p_\tau, p_z, p_{c1})\approx(0.0116313,0.018695, 0.0194174, 0.0235228),
\end{eqnarray}
when taking $w_2=-0.14$.

\begin{figure}[htb]
  \includegraphics[width=0.4\textwidth]{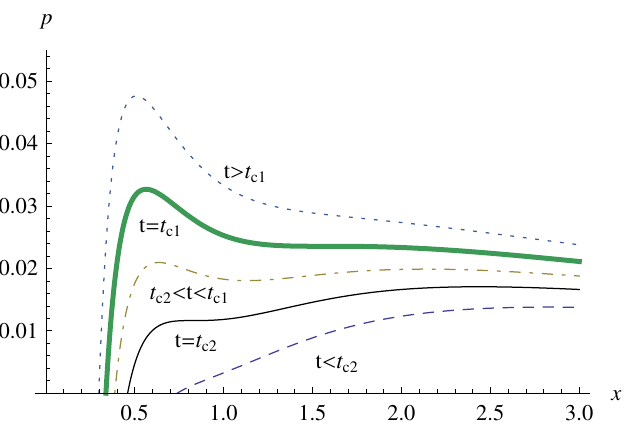}
\hfill%
  \includegraphics[width=0.4\textwidth]{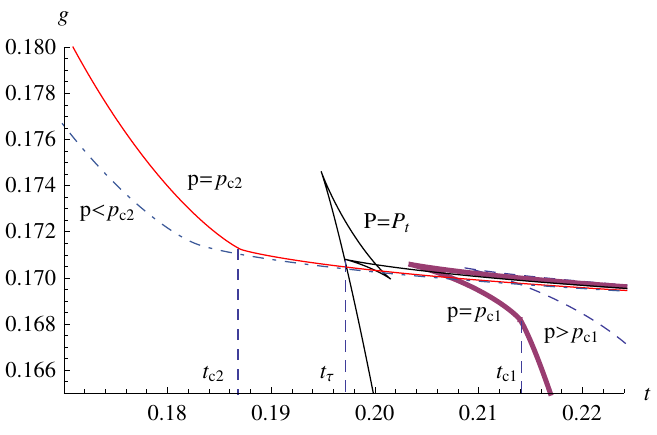}
  \caption{Born-Infeld AdS black holes for $d=4$ and $w_2=-0.14$.
There are two critical points at positive pressure.}\label{fig4}
\end{figure}

\end{enumerate}

With regard to $\left\lvert w_2\right\rvert>\frac{1}{2\pi}$,
the solution of Eq.~(\ref{Frc}) is given by
\begin{eqnarray}
y_3=-\cosh\left(\frac{1}{3}\arccos\left(-2\pi w_2\right)\right),\label{root2}
\end{eqnarray}
which violates the constraint condition $\left\lvert y\right\rvert\leq \frac{1}{\sqrt{2}}$.

All in all, when the parameter $w_2$
satisfies $-\frac{1}{2\pi}<w_2<0$, the Van der Waals-like SBH/LBH phase transition appears.
In addition, the interesting RPT happens in case of $-\frac{1}{2\pi}<w_2<-\frac{1}{\sqrt{8}\pi}$.

\subsection{$P-V$ criticality for $d=5$}

Then Eq.~(\ref{crit1}) can be rewritten as
\begin{eqnarray}
F(x_c)=\frac{1}{x_c}\left(x_c^6+6\right)^{-3/2}
-\frac{5}{18x_c}\left(x_c^6+6\right)^{-1/2}
-\frac{2\pi}{27}\left(w_2+\frac{3w_3}{x_c}\right)=0.\label{PV5}
\end{eqnarray}
Evidently, it is not possible to obtain analytic solution of above equation.
To see more closely the phase transition of the Born-Infeld AdS black hole,
here we analyze the asymptotic property of the function $F(x_c)$.
In addition, the function $\frac{dF(x_c)}{dx_c}$ reads
\begin{eqnarray}
\frac{dF(x_c)}{dx_c}=72\left(1-\frac{5x_c^6}{24}\right)^2+
\frac{135x_c^{12}}{8}
+4\pi w_3\left(x_c^6+6\right)^{5/2}.\label{PVd5}
\end{eqnarray}
Evidently, Eq.~(\ref{PV5}) has more than one real roots.
For different values of $w_2$ and $w_3$, we will investigate the phase
structure and criticality in the extended phase space.

\subsubsection{$w_2>0$ and $w_3>0$}

When $x_c\rightarrow+\infty$, $F(x_c)$ equals to $-\frac{2\pi w_2}{27}$.
Near the origin $x=0$, we have
\begin{eqnarray}
F(x_c)=-\frac{\left(\sqrt{6}+12\pi w_3\right)}{54x_c},\label{Frc2}
\end{eqnarray}
namely,
$F(x_c)$ approaches $-\infty$ on account of $w_3>-\frac{1}{2\sqrt{6}\pi}$.
Moreover, function $\frac{dF(x_c)}{dx_c}$ is always positive,
so there is no a real solution for $x_+$. Therefore, there is no a critical point.

\subsubsection{$w_2<0$ and $w_3>0$}

\begin{figure}[htb]
\centering
\includegraphics[width=0.3\textwidth]{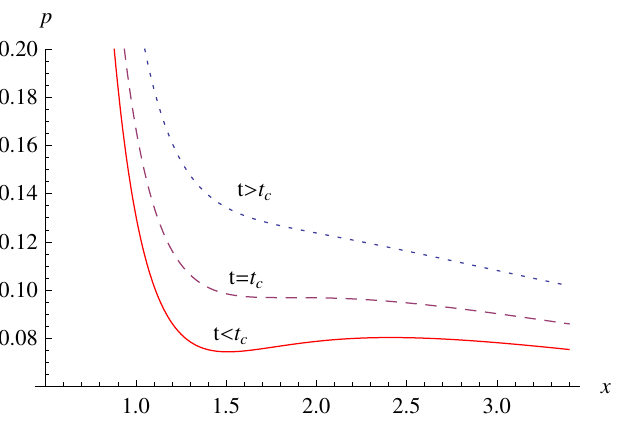}
\hfill%
\includegraphics[width=0.3\textwidth]{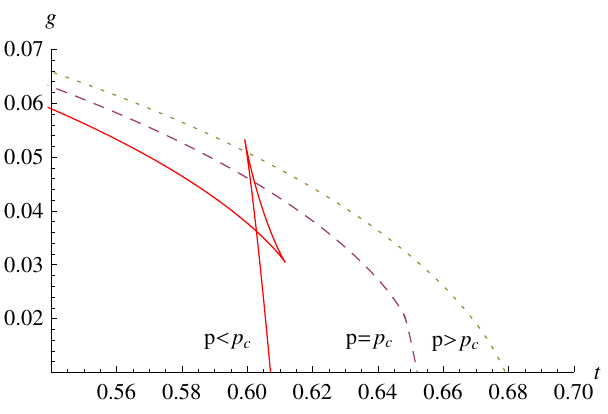}
\hfill%
\includegraphics[width=0.3\textwidth]{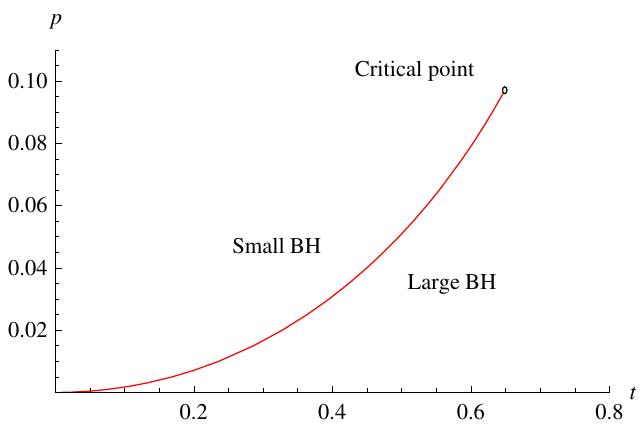}
\caption{Born-Infeld AdS black holes for $d=5$, $w_2=-0.25$ and $w_3=0.1$.
Left: the $p-x$ diagram. The dashed curve represents critical isotherm at $t=t_c$.
The dotted and solid curves correspond to $t>t_c$ and $t<t_c$, respectively.
Center: the $g-t$ diagram. The solid curve represents $p<p_c$, the dotted curve
correspond to $p>p_c$ and the dashed curve is for $p=p_c$. We observe standard swallowtail
behavior. Right: The $p-t$ diagram, showing the coexistence line of SBH/LBH
phase transition terminating at a critical point. These plots are analogous to typical
behavior of the liquid-gas phase transition of the VdW'fluid.}\label{fig5}
\end{figure}

Here we adopt similar discussions above. The function $F(x_c)=-\frac{2\pi w_2}{27}>0$
in case of $w_2<0$. However, $F(x_c)$ approaches $-\infty$ near the origin $x=0$.
Evidently, there is only one positive root of Eq.~(\ref{PV5}) on account
of $\frac{dF(x_c)}{dx_c}>0$. Then, a critical point occurs. In Fig.~{\ref{fig5}},
we display VdW-like small/large black hole phase transition in the system.

\subsubsection{$w_2>0$ and $w_3<0$}

\begin{figure}[htb]
\centering
\subfigure[]{\label{fig:a} 
  \includegraphics[width=0.3\textwidth]{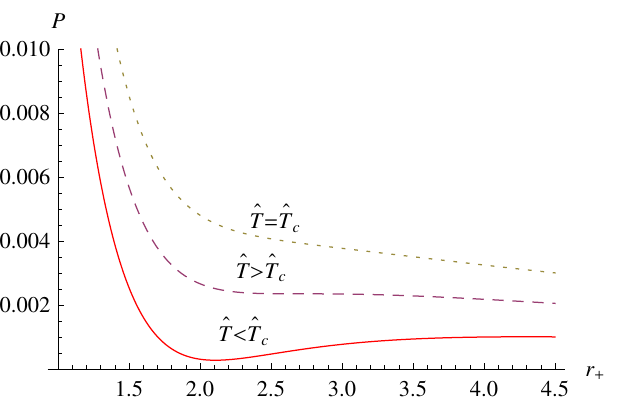}}%
\hfill%
\subfigure[]{\label{fig:b} 
  \includegraphics[width=0.3\textwidth]{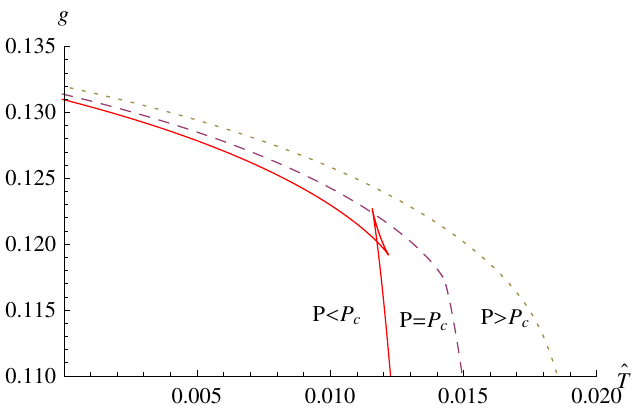}}%
  \hfill%
\subfigure[]{\label{fig:c} 
  \includegraphics[width=0.3\textwidth]{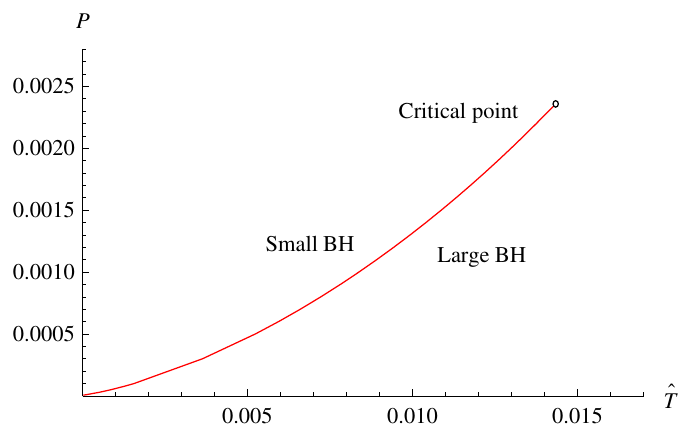}}%
  \caption{Born-Infeld AdS black holes for $d=5$, $w_2=0.001$ and $w_3=-0.02$.}\label{fig6}
\end{figure}

In this case, it is a hard work to discuss the asymptotic property of Eq.~(\ref{PV5}).
Here we resort to graphical and numerical methods, and also find the existence of
VdW-like small/large black hole phase transition in the system, see Figure.~{\ref{fig6}}.

\subsubsection{$w_2<0$ and $w_3<0$}

In Ref \cite{Hendi:2015hoa}, Hendi et al. pointed out that the VdW-like
SBH/LBH phase transition occurs when $w_2<0$ and $w_3<0$. Actually,
there are some other interesting phase transitions.

\begin{figure}[htb]
  \centering
  \includegraphics{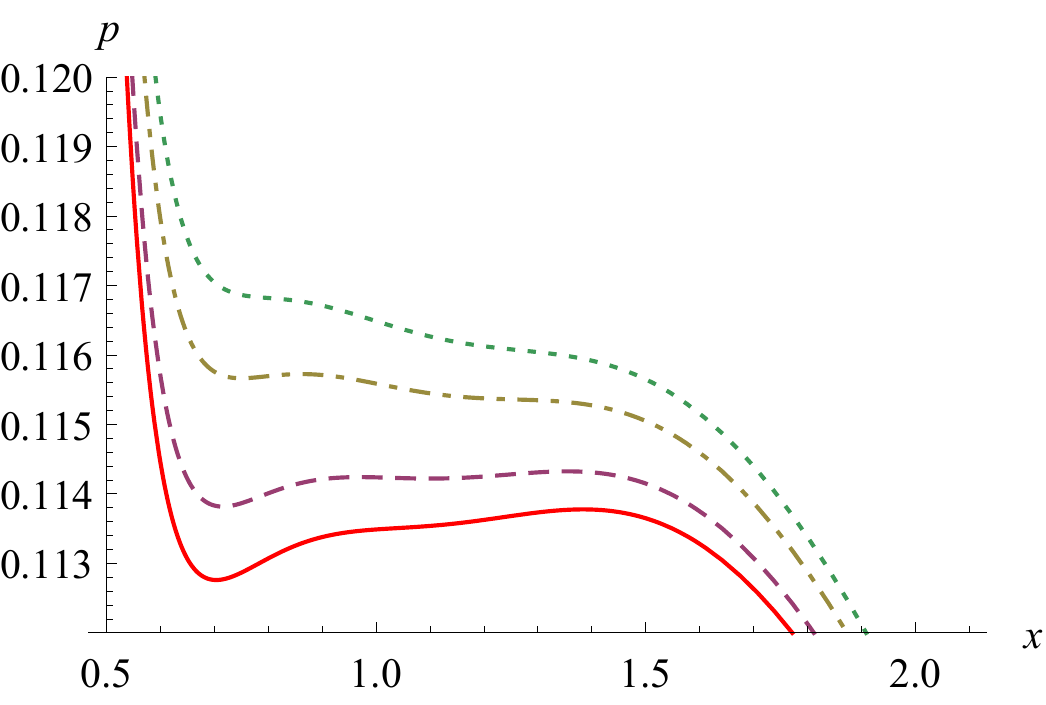}
  \caption{Behavior of $p$ as a function of $x$ for $t=0.495, 0.497, 0.4988$, and 0.5 from
  bottom to top. For small pressure, one can see that there are stable SBH and LBH branches,
  which implies the occurrence of VdW like phase transition.
  With the increasing of the temperature, there appears a
  new stable IBH branch. Further increase the pressure, this branch disappears.}\label{fig7}
\end{figure}

\begin{figure}[htb]
\centering
\subfigure[]{\label{fig:a3} 
  \includegraphics[width=0.3\textwidth]{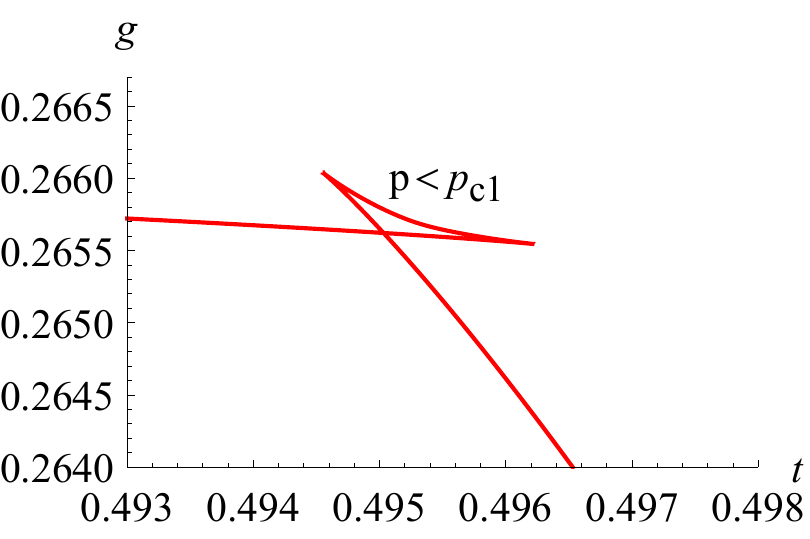}}%
\hfill%
\subfigure[]{\label{fig:b3} 
  \includegraphics[width=0.3\textwidth]{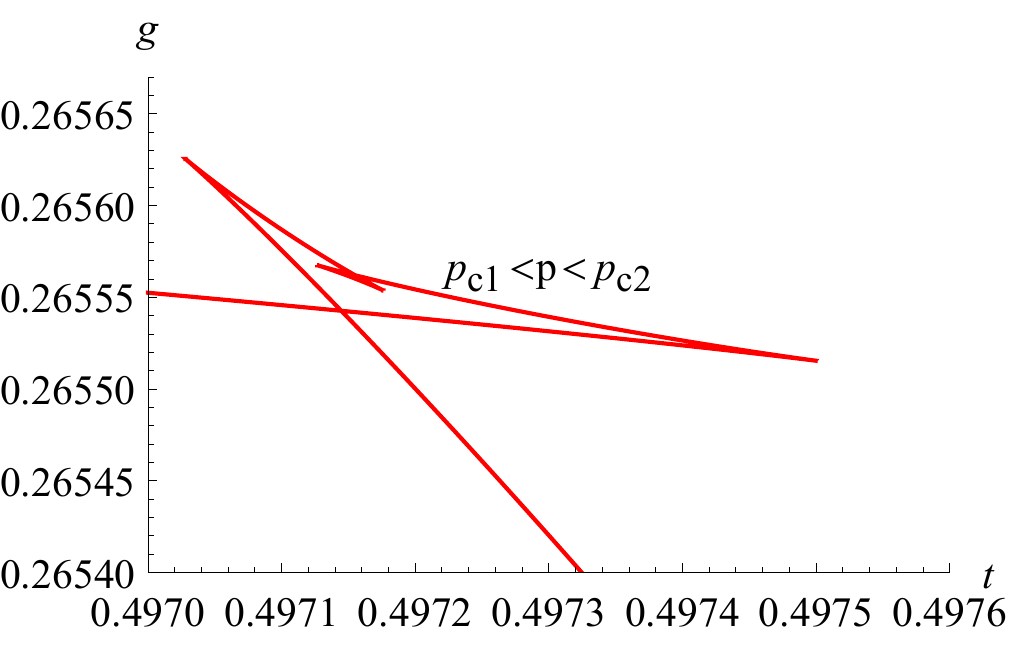}}%
  \hfill%
\subfigure[]{\label{fig:a4} 
  \includegraphics[width=0.3\textwidth]{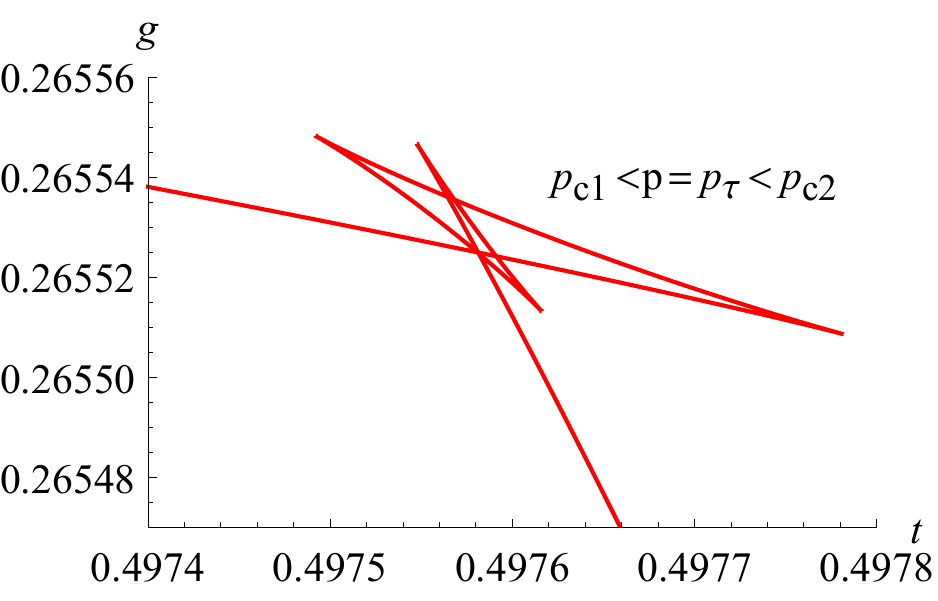}}%
  \hfill%
\subfigure[]{\label{fig:b4} 
  \includegraphics[width=0.3\textwidth]{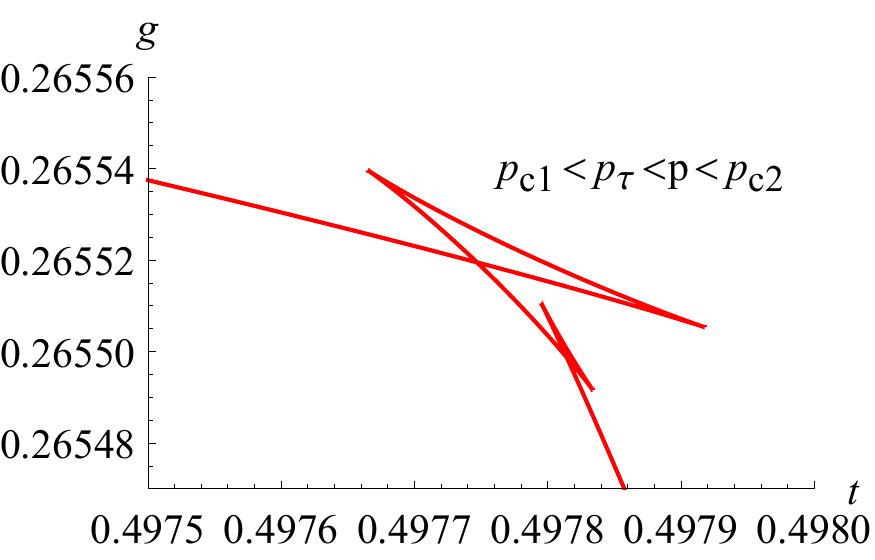}}%
    \hfill%
\subfigure[]{\label{fig:a5} 
  \includegraphics[width=0.3\textwidth]{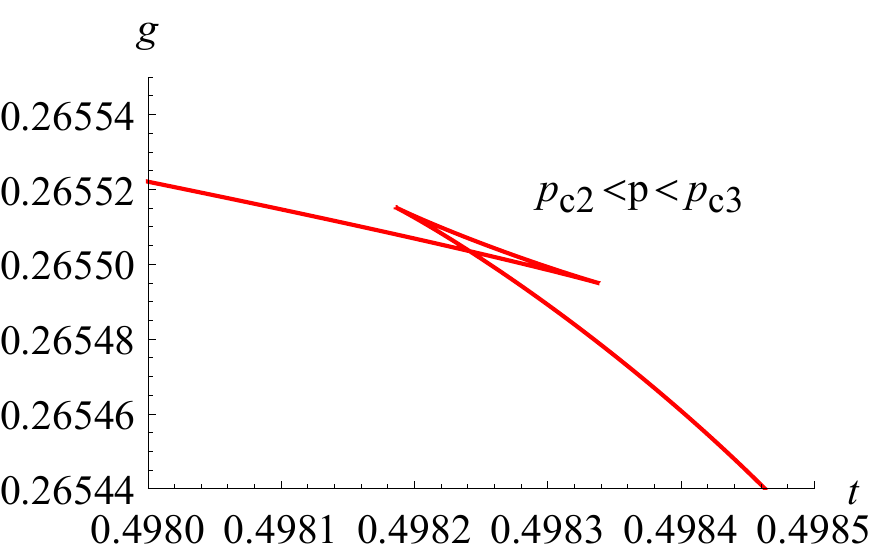}}%
  \hfill%
\subfigure[]{\label{fig:b5} 
  \includegraphics[width=0.3\textwidth]{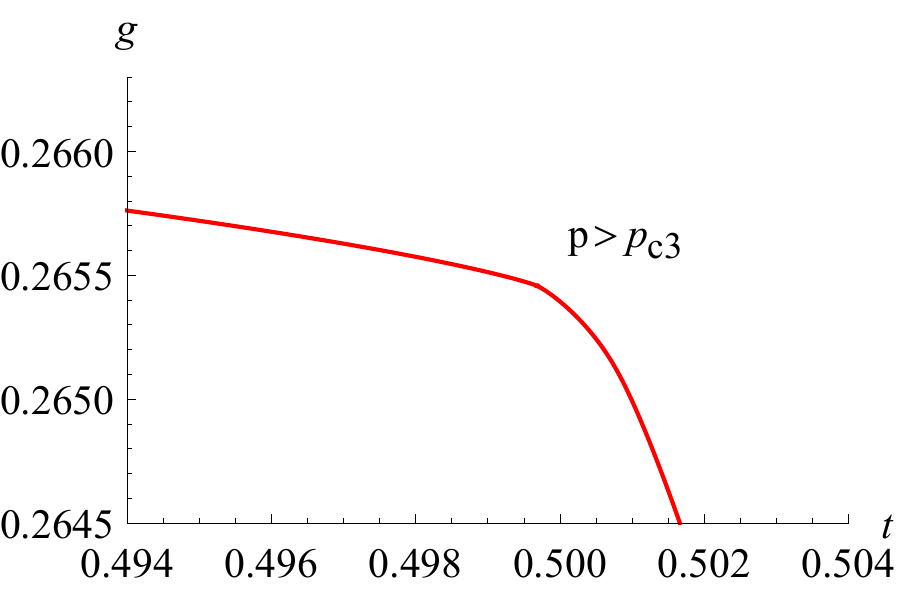}}%
\caption{Gibbs free energy of five dimensional AdS black holes for $w_2=-0.084825$ and $w_3=-0.045$.}\label{fig8}
\end{figure}

For the case of $w_2=-0.084825$ and $w_3=-0.045$,  the pressure $p$ has three critical points,
i.e., $(p_{c1}, p_{c2}, p_{c3})=(0.113983, 0.115076, 0.116354)$
and $(t_{c1}, t_{c2}, t_{c3})=(0.496659, 0.498326, 0.499503)$. We plot the pressure
$p$ as a function of $x$ for $t=0.48, 0.497, 0.4988$, and 0.51
(from bottom to top) in Fig.~(\ref{fig7}).
When $p<p_{c1}$, there exists a characteristic swallow tail behavior in the
$g-t$ diagram, and a VdW-like SBH/LBH phase transition will occur.
Further increasing $p$ such that $p_{c1}<p<p_{c2}$, there appears a new stable
IBH branch. For the corresponding Gibbs free energy in Fig.~(\ref{fig8}), three black
hole phases (i.e., small, large and intermediate black holes) coexist together. Therefore,
we observe a triple point characterized by $(p_\tau, t_\tau)=(0.01960, 0.11226)$.
Slight above this pressure, the system will emerge a standard SBH/IBH/LBH phase transition
with the increase of $t$ . And such phase transition disappears
when $p_{c2}$ is approached.

Further increasing $p$, the stable IBH branch vanishes
in case of $p_{c2}<p<p_{c3}$. And only one stable
branch survives when $p>p_{c3}$. In the ranges $p<p_{c1}$ and $p_{c2}<p<p_{c3}$, it displays
one characteristic swallow tail behavior in Fig.~(\ref{fig8}).
When $p>p_{c3}$, there is no such behavior.

\section{Conclusions and discussions}
\label{4s}

In the extended phase space, we have studied the phase transition and critical behavior
of topological AdS black holes in the four and five dimensional Born-Infeld-massive gravity.
For $d=4$, we found that when the horizon topology is spherical $(k=1)$,
Ricci flat $(k=0)$ or hyperbolic $(k=-1)$, there always exist the Van der Waals-like
SBH/LBH phase transition when the coupling coefficients of massive potential
is located in the region $-\frac{1}{2\pi}<w_2<0$.
In addition, a monotonic lowering of the temperature yields a large-small-large black
hole transition in the region $-\frac{1}{2\pi}<w_2<-\frac{1}{\sqrt{8}\pi}$,
where we refer to the former large state as an intermediate
black hole (IBH), which is reminiscent of RPTs.
Moreover, this process is also accompanied by a discontinuity
in the global minimum of the Gibbs free energy, referred
to as a zeroth-order phase transition.

In some range of the parameters, there are three critical points for five-dimensional
Born-Infeld AdS black hole. In such range, the Gibbs free energy displays the behavior
of two swallow tails. This phenomenon has been never recovered before.

Recent observations of gravitational waves have put an upper bound of $1.2\times 10^{-22} eV/c^2$ on the graviton's mass\cite{Abbott:2016blz}. We can find in 4-dimensional case, the interesting RPTs can always appear as long as the parameters $q$ and $b$ take the suitable values with the constant $k$ takes the values $\pm1$. When the constant $k=0$, the role of the graviton's mass is highlighted, the parameters $q$ and $b$ cannot take an acceptable range (means in the framework of the Born-Infeld theory) to make the parameter $w_2\in(-\frac{1}{2\pi}, -\frac{1}{\sqrt{8}\pi})$, which means only the VdW-like phase transitions might happen. In the 5-dimensional case, when the constant $k=1$, this interesting phenomenon could appear as long as the parameters $q$ and $b$ take the suitable values. There is no three critical points when the constant $k$ takes $-1$ or $0$, because the parameter $w_2$ is always positive.

Ref.\cite{Zou:2013owa} shows that the RPTs only exist in the 4-dimensional
Born-Infeld AdS black hole with a spherical horizon , and also gives the proof
that there is no reentrant phase transition in the system of higher($\ge 5$) dimensional Born-Infeld AdS black hole.
Ref.\cite{Xu:2015rfa} demonstrated that there only exist the Van der Waals like phase transition in the 4-dimensional AdS black hole in massive gravity with Maxwell's electromagnetic field theory. Our results reveal that the nonlinear electromagetic field plays an important role in the phase transition of the 4-dimensional AdS black hole, and the massive gravity could bring richer phase structures and critical behavior (triple critical points) than that of the Born-Infeld term in the 5-dimensional AdS black hole.

Recently, the charged black hole \cite{Hendi:2015pda},
Born-Infeld black hole \cite{Hendi:2016yof}, and black hole in the
Maxwell and Yang-Mills fields \cite{Meng:2016its}
have been constructed in Gauss-Bonnet-massive gravity.
Only Van der Waals like first order SBH/LBH phase transition exist in these models.
In addition, this RPT and triple points also occur in the higher-dimensional rotating AdS
black holes \cite{Altamirano:2013ane,Altamirano:2014tva}, and higher-dimensional
Gauss-Bonnet AdS black hole \cite{Wei:2014hba,Frassino:2014pha,Hennigar:2016ekz}.
It would be interesting to extend our discussion to these black holes
in Gauss-Bonnet and and 3rd-order Lovelock-massive gravity.


{\bf Acknowledgments}

The work is supported by the National Natural Science Foundation of China under Grant Nos.11647050, 11605152, 11675139 and 51575420, Scientific Research Program Funded by Shaanxi Provincial Education Department under Program No.16JK1394, and Natural Science Foundation of Jiangsu Province under Grant No.BK20160452.

\end{document}